\documentclass[lettersize,journal]{IEEEtran}
\usepackage{amsmath,amsfonts}
\usepackage{algorithmic}
\usepackage{array}
\usepackage[caption=false,font=normalsize,labelfont=sf,textfont=sf]{subfig}
\usepackage{textcomp}
\usepackage{stfloats}
\usepackage{url}
\usepackage{verbatim}
\usepackage{graphicx}
\def\BibTeX{{\rm B\kern-.05em{\sc i\kern-.025em b}\kern-.08em
    T\kern-.1667em\lower.7ex\hbox{E}\kern-.125emX}}
\usepackage{balance}

\usepackage{threeparttable}
\usepackage{enumitem}            
\usepackage{amssymb}            
\usepackage{multirow}            
\usepackage{booktabs}            
\usepackage[ruled]{algorithm2e}  
\usepackage{caption}            
\usepackage{array}
\usepackage[caption=false,font=normalsize,labelfont=sf,textfont=sf]{subfig}
\usepackage{booktabs}
\usepackage{makecell}
\usepackage{multirow}
\usepackage[table]{xcolor}
\definecolor{convbg}{RGB}{242,246,250}
\definecolor{presbg}{RGB}{249,245,238}
\usepackage{tabularx}
\usepackage{array}
\usepackage{booktabs}

\begin{document}
\title{Rethinking Backdoor Repair Evaluation: Distinguishing Aggregate Clean Utility from Benign Performance Preservation}

\author{\IEEEauthorblockN{Baogang Song,
Changtian Song, 
Jian Chen,
Fan He,
Junwei Zhou,
Jianwen Xiang} and
Dongdong Zhao\thanks{
This work was supported by the Hubei Province Major Science and Technology Innovation Program under Grant 2024BAA011. (Corresponding author: Dongdong Zhao, e-mail: zdd@whut.edu.cn.)}

\IEEEauthorblockA{School of Artificial Intelligence, Wuhan University of Technology}}


\maketitle

\begin{abstract}
Backdoor repair aims to suppress malicious behavior in compromised models while preserving their benign task performance. Existing studies typically evaluate these two objectives using Attack Success Rate (ASR) and Overall Clean Accuracy. However, Overall Clean Accuracy aggregates performance across classes and may therefore obscure substantial degradation concentrated in a small portion of the label space. As a result, a repaired model can exhibit suppressed ASR and nearly unchanged aggregate clean accuracy while still losing considerable clean performance on specific classes. In this work, we revisit benign-performance evaluation in backdoor repair from a preservation perspective. We distinguish aggregate clean utility from the preservation of previously available class-wise performance and define class-wise preservation loss by comparing clean performance before and after repair. We further show that aggregation can obscure localized degradation through both localized-loss dilution and cross-class compensation. To complement Overall Clean Accuracy, we characterize localized preservation loss from two complementary perspectives: Worst-Class Preservation Loss, which captures the most severe class-wise degradation, and Tail Preservation Loss, which summarizes the upper tail of the class-wise loss distribution. We conduct a systematic empirical study across representative backdoor attacks, repair methods, datasets, attack targets, and model architectures, with additional validation under clean-label attacks. Our results show that effective attack suppression and favorable aggregate clean performance do not necessarily imply that previously available benign performance is uniformly preserved across classes. Substantial localized preservation losses can remain, and their severity and class-wise structure vary across repair conditions. These findings suggest that aggregate clean utility and localized benign-performance preservation capture different aspects of repair quality, motivating preservation-oriented class-wise evaluation alongside ASR and Overall Clean Accuracy.
\end{abstract}


\section{Introduction}

Backdoor attacks implant hidden malicious behaviors into deep neural networks, allowing a compromised model to maintain strong predictive performance on benign inputs while producing attacker-specified outputs when a particular trigger condition is present~\cite{gu2017badnets,liu2018trojaning}. For backdoored models that have already been trained or deployed, retraining a trusted model from scratch may be impractical because of limited access to the original training data, computational cost, or deployment constraints. Consequently, substantial research has focused on post-training defenses that directly modify an existing compromised model, typically using a limited amount of trusted data~\cite{liu2018finepruning,li2021nad,wu2021anp}. In this work, we refer to this class of model-modifying post-training defenses as \emph{backdoor repair}. Across these methods, suppressing malicious behavior while maintaining benign task performance is a central objective. Accordingly, backdoor repair is commonly evaluated along these two dimensions: Attack Success Rate (ASR) characterizes attack suppression, whereas Overall Clean Accuracy characterizes aggregate benign task performance~\cite{wu2025backdoorbench}.

\begin{figure}[t]
    \centering
    \includegraphics[width=\linewidth]{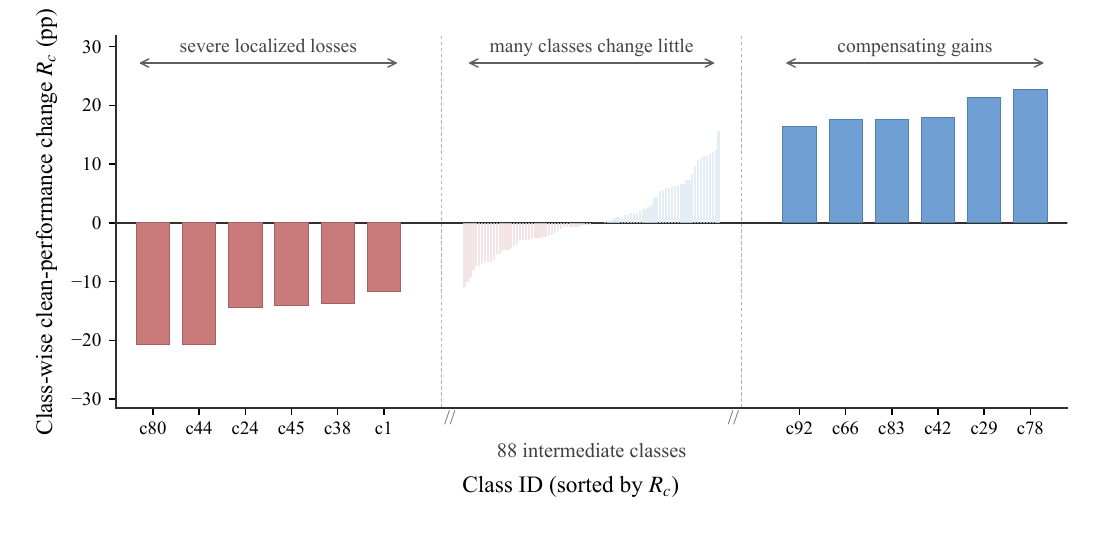}
    \caption{Class-wise clean-performance changes for a representative CIFAR-100 repair condition using D3 on an Input-Aware backdoored model, averaged over three repair repetitions. Despite a mean ASR of 0.05\% and a mean 0.72 pp improvement in Overall Clean Accuracy, the worst-class clean-accuracy loss ranges from 24 to 26 pp across repetitions.}
    \label{fig:motivation}
\end{figure}

Overall Clean Accuracy provides a natural measure of the repaired model’s aggregate clean utility. In this work, benign-performance preservation specifically refers to the retention of pre-repair clean performance at the class level. This raises a related but distinct evaluation question: whether the class-wise clean performance available before repair is retained after repair. A repaired model may exhibit a favorable aggregate clean outcome while individual classes undergo markedly different changes relative to their pre-repair performance. Thus, aggregate clean utility does not by itself establish how well previously available benign performance is preserved across classes. This distinction motivates the central question of this work: When a backdoor repair outcome exhibits effective attack suppression and favorable Overall Clean Accuracy, is this aggregate metric sufficient to characterize the preservation of pre-repair benign performance across classes?

The limitation of aggregate evaluation for assessing benign-performance preservation becomes apparent in a seemingly favorable repair outcome. Figure~\ref{fig:motivation} shows a representative case on CIFAR-100, where D3~\cite{wei2025d3} reduces the ASR of an Input-Aware~\cite{nguyen2020inputaware} backdoored model to 0.05\% while slightly improving Overall Clean Accuracy by 0.72 pp. Viewed only through these conventional metrics, the repair appears to achieve effective attack suppression while maintaining favorable aggregate clean performance. However, a class-wise examination reveals substantial variation that is not apparent from the aggregate result. Some classes exhibit pronounced performance losses, while others improve substantially. These opposing changes can partially offset one another at the aggregate level, while losses confined to a small subset of classes may have only a limited effect on Overall Clean Accuracy. Consequently, a favorable aggregate outcome can coexist with substantial loss of previously available performance on particular classes. This example makes the distinction concrete: aggregate clean utility summarizes the repaired model’s overall clean performance, whereas benign-performance preservation concerns whether the performance available before repair is retained across classes.

Prior work has shown that backdoor mitigation can disproportionately impair the attack-target class~\cite{hsieh2024activation}, providing an important example of localized benign degradation after repair. Such observations indicate that repair-induced benign changes can be highly class dependent and may therefore require examination beyond aggregate clean performance. We characterize these changes from a preservation-oriented perspective by comparing the clean performance of each class before and after repair, using the pre-repair model as the natural reference. This comparison preserves the beneficial interpretation of performance gains while explicitly identifying losses of previously available class-wise benign performance across the label space.

Under this preservation-oriented view, the full set of class-wise changes retains information that aggregate evaluation compresses into a single outcome, allowing performance gains to be distinguished from losses of previously available class-wise benign performance. Because the number of class-wise preservation losses scales with the size of the label space and a complete profile is produced for every repair condition, directly comparing these profiles across attacks, repair methods, and datasets becomes cumbersome. We therefore complement the full profile with two compact summaries: the most severe class-wise loss and the upper tail of the class-wise preservation-loss distribution. These preservation summaries complement rather than replace conventional evaluation: ASR characterizes attack suppression, Overall Clean Accuracy characterizes aggregate clean utility, and the preservation-oriented analysis characterizes localized loss of previously available benign performance.

We conduct a systematic empirical study across representative attacks, repair methods, datasets, attack targets, and model architectures, with additional validation under clean-label attacks. Across these settings, we repeatedly find that effective attack suppression and favorable aggregate clean performance can coexist with substantial localized preservation loss. The affected classes are not limited to the attack target, nor do they form a fixed vulnerable subset across repair conditions. Together, these findings reveal a benign-side blind spot in conventional backdoor-repair evaluation: aggregate clean performance can remain favorable while substantial losses of previously available class-wise benign performance remain hidden. As a result, conventional metrics may provide an incomplete account of the benign consequences of a repair and may fail to distinguish repair outcomes with similar aggregate utility but substantially different preservation losses.

Our main contributions are as follows:

\begin{itemize}

\item We identify a benign-side blind spot in conventional backdoor-repair evaluation by distinguishing aggregate clean utility from the preservation of previously available benign performance. Overall Clean Accuracy characterizes the former but does not by itself establish the latter across classes.

\item We operationalize this preservation-oriented perspective by characterizing class-wise benign performance changes relative to the pre-repair model and separating performance gains from preservation losses. We retain the complete class-wise loss profile and complement it with compact worst-class and upper-tail summaries for systematic comparison.

\item We conduct a systematic empirical study across representative attacks, repair methods, datasets, attack targets, and model architectures, with additional validation under clean-label attacks. We repeatedly observe that effective attack suppression and favorable aggregate clean performance can coexist with substantial localized preservation loss. We further find that these losses are neither confined to the attack-target class nor associated with a fixed vulnerable subset of classes, but vary across repair procedures and experimental conditions.

\end{itemize}
\section{Related Work}
\label{sec:related_work}

\subsection{Post-Training Backdoor Defenses}
\label{sec:related_repair}

Backdoor defenses have been developed at different stages of the model learning and deployment pipeline. Representative directions include screening training data for poisoned samples~\cite{tran2018spectral,chen2019activation}, detecting suspicious inputs or trigger-induced behavior at inference time~\cite{gao2019strip,chou2020sentinet}, and inspecting or modifying trained models after training~\cite{wang2019neuralcleanse,liu2018finepruning}. Among these directions, post-training model-modifying defenses operate directly on already trained compromised models and aim to mitigate backdoor behavior without requiring control over the original model-training process~\cite{lin2024tsbd,wei2025d3}. These methods typically modify model parameters, structures, or internal representations, often using a limited amount of trusted clean data. We focus on this class of defenses, which we refer to as \emph{backdoor repair}.

Early post-training defenses explored trigger reverse engineering and neuron pruning as mechanisms for mitigating backdoor behavior. Neural Cleanse reverse-engineers candidate trigger patterns from a trained model and uses the recovered information to support mitigation~\cite{wang2019neuralcleanse}, whereas Fine-Pruning removes neurons that are weakly activated by clean inputs and subsequently fine-tunes the pruned model~\cite{liu2018finepruning}. Subsequent methods more directly target backdoor-sensitive model components or representations. ANP identifies neurons that are vulnerable to adversarial perturbations and prunes them~\cite{wu2021anp}; AWM jointly estimates adversarial trigger-related perturbations and learns masks over sensitive weights through min--max optimization~\cite{chai2022awm}; and RNP uses an unlearning--recovery procedure to expose and prune backdoor-related neurons~\cite{li2023rnp}. Moving beyond explicit pruning or masking, Neural Polarizer inserts a lightweight learnable transformation to suppress trigger-related representations while preserving benign information~\cite{zhu2023polarizer}.

Beyond explicitly identifying or suppressing backdoor-sensitive components, later repair methods have increasingly relied on knowledge transfer, unlearning, and specialized optimization objectives to update compromised models. NAD transfers attention patterns from a cleanly fine-tuned teacher to the backdoored model~\cite{li2021nad}. I-BAU formulates backdoor mitigation as a minimax unlearning problem optimized through implicit hypergradients~\cite{zeng2022ibau}, while SAU mitigates backdoor behavior by unlearning shared adversarial examples associated with backdoor risk~\cite{wei2023sau}. Fine-tuning-based repair has also evolved beyond standard optimization: FT-SAM incorporates sharpness-aware minimization into the fine-tuning process~\cite{zhu2023ftsam}, whereas FST promotes feature shifts by moving classifier weights away from their compromised states~\cite{min2023fst}. TSBD further uses weight changes induced by clean unlearning to identify backdoor-related parameters, followed by reinitialization and activeness-aware fine-tuning~\cite{lin2024tsbd}.

Recent repair methods continue to explore different strategies for modifying compromised models. D3 encourages the repaired model to move away from the neighborhood of its initial backdoored weights through distance-driven constrained optimization~\cite{wei2025d3}. PGBD performs post-hoc sanitization by exploiting activation-space geometry to penalize movements toward the attack target~\cite{amula2025pgbd}, while UPGP combines unlearning perturbation with orthogonality-constrained gradient projection to suppress backdoor behavior while limiting interference with the model's main task~\cite{li2026upgp}. Despite their methodological differences, these repair methods broadly share the goal of suppressing malicious behavior while maintaining benign predictive utility. Rather than developing another repair method, our work focuses on how the benign performance of repaired models is evaluated.

\subsection{Evaluation of Backdoor Defenses}
\label{sec:related_evaluation}

The growing diversity of backdoor attacks and defenses has motivated increasing efforts toward more systematic and reproducible evaluation. BackdoorBench provides a unified implementation framework and standardized protocol for comparing representative attacks and defenses across datasets and model architectures~\cite{wu2022backdoorbench}. Its expanded version further broadens the coverage of attacks, defenses, architectures, and analysis tools, while standardizing evaluation with metrics such as Clean Accuracy (C-Acc), Attack Success Rate (ASR), Robust Accuracy (R-Acc), and composite measures including Defense Effectiveness Rate (DER)~\cite{wu2025backdoorbench}. Beyond benchmarking, the SoK by Abad et al.~\cite{abad2025sok} documents substantial heterogeneity across the backdoor-defense literature in experimental protocols, evaluation metrics, threat-model assumptions, and hyperparameter selection. Recent work has further examined assumptions embedded in purification protocols themselves; for example, Wei et al.~\cite{wei2025auxiliary} show that purification effectiveness can vary substantially with the source and distribution of the auxiliary clean data available to the defender. Together, these studies show that backdoor-defense evaluation depends not only on the mitigation algorithm itself, but also on the metrics, protocols, and assumptions under which repair outcomes are assessed.

A separate line of work questions whether favorable conventional metrics fully characterize the security state of a repaired model. Min et al.~\cite{min2024superficial} show that models with low post-purification ASR can remain susceptible to rapid recovery of backdoor behavior, motivating post-purification robustness as an additional security consideration. Zhu et al.~\cite{zhu2024reactivation} similarly demonstrate that backdoors suppressed by existing post-training defenses can remain dormant and later be reactivated. Together, these studies reveal a limitation on the \emph{security} side of conventional evaluation: a low observed ASR does not necessarily imply that the backdoor has been robustly removed.

Benign effects of repair have received less systematic attention from an evaluation perspective, although several studies report relevant class-specific phenomena. Hsieh et al.~\cite{hsieh2024activation} show that backdoor unlearning can disproportionately impair the attack-target class and introduce an additional repair step to recover its utility. Related class-wise effects have also been studied in broader neural-network repair settings. Chen et al.~\cite{chen2024inner} examine repair-induced inter-class accuracy imbalance from a fairness perspective. These studies provide important evidence that repair can affect classes unevenly. However, their primary objectives are to characterize particular forms of class-specific degradation or disparity, rather than to systematically evaluate how well the benign performance available before repair is preserved across the label space. Our work takes this preservation-oriented perspective and studies repair-induced class-wise performance changes relative to the pre-repair model as a distinct aspect of backdoor-repair evaluation.

Despite these class-specific observations, contemporary post-training defenses still commonly characterize benign utility through overall task performance or clean accuracy while reporting backdoor suppression~\cite{amula2025pgbd,li2026upgp}. Such aggregate measures characterize overall clean utility but do not by themselves establish how much of each class's pre-repair clean performance is retained after repair. Consequently, the preservation of previously available benign performance remains insufficiently characterized as a distinct dimension of backdoor-repair evaluation. Our work addresses this gap by treating attack suppression, aggregate clean utility, and localized benign-performance preservation as distinct but complementary evaluation dimensions. This perspective extends conventional backdoor-repair evaluation beyond aggregate benign performance by explicitly characterizing how well pre-repair class-wise benign performance is preserved across the class space, thereby providing a more complete account of the benign consequences of repair.
\section{Beyond Aggregate Evaluation of Benign Performance}
\label{sec:preservation}

Benign-side evaluation of backdoor repair raises two related but distinct questions: how well the repaired model performs overall, and how much of the class-wise benign performance available before repair is retained. Overall Clean Accuracy provides a natural measure of the former, whereas the latter requires examining repair-induced performance changes relative to the pre-repair state. We therefore characterize the clean-performance change of each class before and after repair, treating performance improvements as beneficial changes and decreases from pre-repair performance as preservation losses. This section formalizes this preservation-oriented perspective by defining signed class-wise performance change and preservation loss, analyzing their relation to aggregate clean utility, and introducing compact summaries of localized preservation loss.

\subsection{Class-Wise Benign-Performance Change and Preservation Loss}
\label{sec:preservation_definition}

To characterize benign-performance preservation at the class level, we use the pre-repair model as the reference, so that each class is evaluated against the clean performance available immediately before repair. Let $f^{\mathrm{pre}}$ and $f^{\mathrm{rep}}$ denote the models before and after repair, respectively, and let $A_c(f)$ denote the clean accuracy of model $f$ on class $c$. We define the signed class-wise performance change as
\begin{equation}
R_c
=
A_c(f^{\mathrm{rep}})
-
A_c(f^{\mathrm{pre}}).
\label{eq:classwise_change}
\end{equation}
Here, $R_c$ captures both the direction and magnitude of the clean-performance change for class $c$: $R_c>0$ indicates an improvement after repair, whereas $R_c<0$ indicates a degradation relative to its pre-repair performance.

From a preservation perspective, the relevant quantity is the decrease, if any, from each class's pre-repair performance. We therefore define the \emph{class-wise preservation loss} as
\begin{equation}
D_c
=
[-R_c]_+
=
\left[
A_c(f^{\mathrm{pre}})
-
A_c(f^{\mathrm{rep}})
\right]_+,
\label{eq:classwise_preservation_loss}
\end{equation}
where $[x]_+=\max(x,0)$. Thus, $D_c=0$ when the clean performance of class $c$ is preserved or improved, whereas $D_c>0$ quantifies the amount of previously available benign performance that is lost after repair.

The signed change $R_c$ retains both the direction and magnitude of the class-wise clean-performance change, whereas $D_c$ isolates the downside relevant to benign-performance preservation. For a task with $C$ classes, the collection $\{D_c\}_{c=1}^{C}$ forms the complete class-wise preservation-loss profile, retaining information about both where preservation loss occurs and how severe it is. This full profile serves as the primary representation of preservation loss in our analysis. The compact summaries introduced in Section~\ref{sec:preservation_characterization} provide complementary views for systematic comparison across repair conditions.

\subsection{Aggregate Clean Utility and Benign-Performance Preservation}
\label{sec:aggregate_preservation}

Overall Clean Accuracy characterizes aggregate clean utility. To relate this aggregate view to the class-wise changes defined above, we consider its change from before to after repair. Let $n_c$ denote the number of evaluation samples from class $c$, let $N=\sum_{c=1}^{C}n_c$, and define $w_c=n_c/N$. The repair-induced change in Overall Clean Accuracy can then be written as
\begin{equation}
\Delta A_{\mathrm{overall}}
=
A_{\mathrm{overall}}(f^{\mathrm{rep}})
-
A_{\mathrm{overall}}(f^{\mathrm{pre}})
=
\sum_{c=1}^{C} w_c R_c.
\label{eq:overall_change}
\end{equation}
Thus, the change in aggregate clean performance is the sample-weighted average of the signed class-wise changes. In the empirical results, we denote this quantity by $\Delta\mathrm{ACC}$.

To make the contributions of improvements and preservation losses explicit, let
\begin{equation}
G_c=[R_c]_+
\label{eq:classwise_gain}
\end{equation}
denote the class-wise performance gain after repair. Together with $D_c=[-R_c]_+$, we have
\begin{equation}
R_c=G_c-D_c.
\label{eq:gain_loss_decomposition}
\end{equation}
Substituting this decomposition into Eq.~\eqref{eq:overall_change} gives
\begin{equation}
\Delta A_{\mathrm{overall}}
=
\sum_{c=1}^{C}w_cG_c
-
\sum_{c=1}^{C}w_cD_c.
\label{eq:overall_gain_loss}
\end{equation}
Equation~\eqref{eq:overall_gain_loss} shows that aggregate clean-performance change reflects the net balance between class-wise gains and preservation losses. It therefore does not, by itself, determine the magnitude or distribution of the underlying preservation losses.

Two aggregation effects make this loss of preservation information explicit. The first is \emph{localized-loss dilution}. Consider a class-balanced task with $C$ classes in which one class incurs a preservation loss of $\delta$ while all other classes remain unchanged:
\begin{equation}
\mathbf{R}
=
(-\delta,0,\ldots,0).
\label{eq:localized_loss_example}
\end{equation}
The corresponding aggregate change is
\begin{equation}
\Delta A_{\mathrm{overall}}
=
-\frac{\delta}{C}.
\label{eq:localized_loss_dilution}
\end{equation}
Thus, even without any compensating improvement, a substantial loss confined to one class is attenuated by aggregation and may produce only a small change in Overall Clean Accuracy. Under the class-balanced setting, its contribution to the aggregate result decreases proportionally with $1/C$.

The second effect is \emph{cross-class compensation}. Consider two class-balanced repair outcomes:
\begin{equation}
\mathbf{R}^{(1)}
=
(0,0,\ldots,0),
\qquad
\mathbf{R}^{(2)}
=
(-\delta,+\delta,0,\ldots,0).
\label{eq:compensation_example}
\end{equation}
Both yield $\Delta A_{\mathrm{overall}}=0$, but their preservation outcomes differ. The first contains no preservation loss, whereas the second incurs a loss of magnitude $\delta$ on one class that is exactly offset in the aggregate by an equal gain on another. Hence, identical aggregate changes can correspond to fundamentally different preservation-loss profiles.

More generally, the mapping from the complete set of class-wise changes $\{R_c\}_{c=1}^{C}$ to the scalar $\Delta A_{\mathrm{overall}}$ is many-to-one. Repair outcomes with the same or similar aggregate clean-performance change can therefore exhibit substantially different preservation-loss profiles. Overall Clean Accuracy remains a meaningful measure of aggregate clean utility; the limitation is that aggregate information alone cannot determine where preservation loss occurs or how severe it is across the label space. Characterizing benign-performance preservation therefore requires retaining information beyond the aggregate clean outcome.

\subsection{Characterization of Localized Preservation Loss}
\label{sec:preservation_characterization}

The complete class-wise preservation-loss profile $\{D_c\}_{c=1}^{C}$ retains fine-grained information about where preservation loss occurs across the label space and how severe it is for each affected class. However, each repair condition produces a full class-wise profile, making direct comparison increasingly cumbersome as the number of classes and experimental conditions grows. We therefore retain the full profile as the primary representation of preservation loss and complement it with two compact summaries: the most severe individual loss and the upper tail of the class-wise preservation-loss distribution.

We first define the \emph{Worst-Class Preservation Loss} as
\begin{equation}
D_{\max}
=
\max_{c\in\{1,\ldots,C\}} D_c .
\label{eq:worst_class_loss}
\end{equation}
$D_{\max}$ captures the largest loss of previously available benign performance among all classes. It therefore provides a direct view of the most severe localized degradation, which may have only a limited influence on aggregate clean utility.

The worst-class value alone, however, does not indicate whether severe preservation loss is isolated to a single class or extends across multiple classes. For example, consider the two preservation-loss profiles
\begin{equation}
(30,0,0,\ldots)
\qquad\text{and}\qquad
(30,28,26,24,\ldots).
\label{eq:tail_motivation}
\end{equation}
Both have the same $D_{\max}$, although the second exhibits substantial preservation loss across a broader portion of the label space. We therefore complement the worst-class view with an upper-tail summary.

For a tail level $\alpha\in[0,1)$, we define the \emph{Tail Preservation Loss} using the class-uniform CVaR representation~\cite{rockafellar2002cvar}:
\begin{equation}
D_{\mathrm{tail}}^{(\alpha)}
=
\min_{\eta}
\left[
\eta
+
\frac{1}{(1-\alpha)C}
\sum_{c=1}^{C}
(D_c-\eta)_+
\right],
\label{eq:tail_preservation_loss}
\end{equation}
where $\eta$ is an auxiliary variable in the CVaR representation, and $1-\alpha$ determines the upper-tail mass under the class-uniform empirical distribution. The resulting quantity summarizes preservation loss within the most affected portion of the class space. Whereas $D_{\max}$ reflects a single extreme class, $D_{\mathrm{tail}}^{(\alpha)}$ captures whether substantial preservation loss extends across multiple highly affected classes.

Classes are weighted uniformly in the tail summary because the preservation analysis is intended to characterize loss across the label space without reducing the influence of a class solely because it contains fewer evaluation samples. Overall Clean Accuracy, in contrast, retains its conventional sample-weighted interpretation as a measure of aggregate clean utility.

Taken together, $\{D_c\}_{c=1}^{C}$ provides the complete class-wise preservation-loss profile, while $D_{\max}$ and $D_{\mathrm{tail}}^{(\alpha)}$ provide compact views of its extreme and upper-tail behavior. These summaries complement rather than replace the full profile or conventional backdoor-repair evaluation. ASR characterizes attack suppression, Overall Clean Accuracy characterizes aggregate clean utility, and the preservation analysis characterizes localized loss of previously available benign performance.
\section{Experimental Setup}
\label{sec:experimental_setup}

Our experiments examine whether conventional attack-suppression and aggregate clean-utility criteria are sufficient to characterize benign-performance preservation after repair. We evaluate this question across representative datasets, model architectures, backdoor attacks, and post-training repair methods under a consistent evaluation framework.

For each method-specific candidate-state set, we first select a representative repaired state using only conventional attack-suppression and aggregate clean-utility criteria. Preservation is evaluated only after this representative state has been fixed, relative to the corresponding pre-repair model. This design allows us to examine whether repair outcomes favored by conventional evaluation can still exhibit substantial class-wise preservation loss.

\subsection{Datasets, Models, Attacks, and Repair Methods}
\label{sec:experimental_settings}

Our experiments consider three image-classification benchmarks with different label-space sizes: CIFAR-10 and CIFAR-100~\cite{krizhevsky2009learning}, containing 10 and 100 classes, respectively, and GTSRB~\cite{stallkamp2011gtsrb}, containing 43 classes. The primary benchmark uses ResNet18~\cite{he2016resnet} on CIFAR-10 and CIFAR-100 and PreActResNet18~\cite{he2016identity} on GTSRB. To evaluate whether the observed preservation behavior generalizes across model architectures, we additionally consider PreActResNet18 and VGG19-BN~\cite{simonyan2015vgg} on CIFAR-100 in Section~\ref{sec:generalization}. Dataset preprocessing, pre-repair model construction, and other implementation details are provided in Appendix~\ref{app:pre_repair_attack}.

The primary benchmark includes three backdoor attacks with different trigger constructions: BadNet~\cite{gu2017badnets}, WaNet~\cite{nguyen2021wanet}, and Input-Aware~\cite{nguyen2020inputaware}. BadNet uses a fixed trigger pattern, WaNet generates triggered inputs through spatial warping, and Input-Aware produces input-dependent triggers. All three attacks are instantiated in the all-to-one setting. The primary experiments use class $0$ as the attack target, while target classes $3$ and $6$ are additionally evaluated on CIFAR-100 to examine target generalization. Detailed attack configurations are provided in Appendix~\ref{app:pre_repair_attack}, while the performance of the pre-repair models used in the primary benchmark is reported in Appendix~\ref{app:experimental_protocol}.

The primary benchmark evaluates six post-training repair methods: standard fine-tuning (FT), Fine-Pruning (FP)~\cite{liu2018finepruning}, Neural Attention Distillation (NAD)~\cite{li2021nad}, Adversarial Neuron Pruning (ANP)~\cite{wu2021anp}, I-BAU~\cite{zeng2022ibau}, and D3~\cite{wei2025d3}. These methods cover a range of repair strategies, including clean fine-tuning, pruning-based repair, knowledge distillation, adversarial neuron suppression, adversarial unlearning, and distance-driven optimization. We additionally conduct a separate clean-label validation on CIFAR-10 using LC~\cite{turner2019labelconsistent} and SIG~\cite{barni2019backdoor} as attacks and FST~\cite{min2023fst} and RNP~\cite{li2023rnp} as repair methods. Detailed repair configurations and predefined candidate states are provided in Appendix~\ref{app:repair_details}, and the clean-label results are reported in Section~\ref{sec:clean_label_validation}.

\begin{table*}[t]
    \centering
    \caption{
    Conventional and preservation-oriented evaluation of representative repair states under the primary target-0 setting.
    }
    \label{tab:main_preservation_results}
    \setlength{\tabcolsep}{3.2pt}
    \renewcommand{\arraystretch}{1.08}

    \begin{tabular}{ll*{2}{rrrrrr}rrrr}
        \toprule
        \multirow{2}{*}{Dataset}
        & \multirow{2}{*}{Metric}
        & \multicolumn{6}{c}{BadNet}
        & \multicolumn{6}{c}{WaNet}
        & \multicolumn{4}{c}{Input-Aware} \\
        \cmidrule(lr){3-8}
        \cmidrule(lr){9-14}
        \cmidrule(lr){15-18}

        &
        & FT & FP & NAD & ANP & I-BAU & D3
        & FT & FP & NAD & ANP & I-BAU & D3
        & FP & ANP & I-BAU & D3 \\
        \midrule

        \rowcolor{convbg}
        \cellcolor{white}
        & \cellcolor{white} ASR (\%)
        & 3.00 & 2.23 & 3.76 & 5.07 & 3.57 & 0.64
        & 9.83 & 6.23 & 7.82 & 7.18 & 1.46 & 0.55
        & 7.37 & 2.93 & 1.14 & 0.00 \\

        \rowcolor{convbg}
        \cellcolor{white}
        & \cellcolor{white} $\Delta$ACC (pp)
        & -0.56 & 0.00 & -6.71 & -1.20 & -2.28 & -0.61
        & 0.93 & 0.33 & -2.29 & 1.27 & -1.88 & 0.01
        & 2.14 & 0.09 & -1.11 & 1.87 \\

        \addlinespace[1.5pt]

        \rowcolor{presbg}
        \cellcolor{white}
        & \cellcolor{white} $D_{\mathrm{tail}}^{(.8)}$ (pp)
        & 2.18 & 1.37 & 16.93 & 8.78 & 7.37 & 3.02
        & 0.55 & 1.72 & 5.82 & 1.27 & 7.70 & 2.43
        & 1.62 & 0.65 & 9.40 & 2.20 \\

        \rowcolor{presbg}
        \cellcolor{white}\multirow{-4}{*}{CIFAR-10}
        & \cellcolor{white} $D_{\max}$ (pp)
        & 2.60 & 1.57 & 18.00 & 15.53 & 9.23 & 3.77
        & 0.75 & 2.20 & 6.23 & 2.47 & 8.17 & 2.90
        & 2.17 & 0.93 & 12.40 & 2.63 \\
        \midrule

        \rowcolor{convbg}
        \cellcolor{white}
        & \cellcolor{white} ASR (\%)
        & 1.81 & 3.74 & 0.94 & 5.73 & 0.77 & 2.06
        & 8.99 & 6.03 & 9.45 & 2.27 & 6.79 & 0.30
        & 8.35 & 4.97 & 4.82 & 0.05 \\

        \rowcolor{convbg}
        \cellcolor{white}
        & \cellcolor{white} $\Delta$ACC (pp)
        & -0.58 & -0.55 & -8.98 & -1.91 & -4.60 & -7.72
        & 1.98 & -1.14 & -3.14 & 2.20 & -1.46 & -7.07
        & 5.17 & 0.22 & -0.69 & 0.72 \\

        \addlinespace[1.5pt]

        \rowcolor{presbg}
        \cellcolor{white}
        & \cellcolor{white} $D_{\mathrm{tail}}^{(.8)}$ (pp)
        & 7.05 & 6.53 & 21.98 & 12.43 & 17.25 & 18.22
        & 3.27 & 9.30 & 13.85 & 4.32 & 11.73 & 17.63
        & 3.60 & 6.35 & 15.88 & 11.13 \\

        \rowcolor{presbg}
        \cellcolor{white}\multirow{-4}{*}{CIFAR-100}
        & \cellcolor{white} $D_{\max}$ (pp)
        & 13.67 & 16.00 & 30.33 & 52.00 & 29.00 & 29.67
        & 9.00 & 15.67 & 19.00 & 16.67 & 24.50 & 25.67
        & 9.67 & 11.33 & 29.50 & 25.00 \\
        \midrule

        \rowcolor{convbg}
        \cellcolor{white}
        & \cellcolor{white} ASR (\%)
        & 1.92 & 0.00 & 2.07 & 3.37 & 0.01 & 0.01
        & 2.96 & 5.26 & 1.01 & 1.02 & 0.01 & 0.00
        & 0.39 & 0.06 & 0.85 & 0.00 \\

        \rowcolor{convbg}
        \cellcolor{white}
        & \cellcolor{white} $\Delta$ACC (pp)
        & 0.47 & 0.94 & 0.28 & -0.33 & -1.01 & 0.44
        & 0.87 & 1.34 & 0.78 & 0.39 & 0.74 & 0.97
        & 1.75 & 0.20 & 0.69 & 1.42 \\

        \addlinespace[1.5pt]

        \rowcolor{presbg}
        \cellcolor{white}
        & \cellcolor{white} $D_{\mathrm{tail}}^{(.8)}$ (pp)
        & 1.64 & 2.18 & 3.08 & 2.94 & 10.95 & 4.54
        & 1.77 & 0.94 & 1.55 & 2.72 & 4.25 & 2.39
        & 3.28 & 4.56 & 6.60 & 3.16 \\

        \rowcolor{presbg}
        \cellcolor{white}\multirow{-4}{*}{GTSRB}
        & \cellcolor{white} $D_{\max}$ (pp)
        & 7.52 & 10.74 & 10.19 & 7.12 & 29.44 & 17.22
        & 6.00 & 5.33 & 5.56 & 10.00 & 16.39 & 11.33
        & 14.44 & 14.44 & 25.19 & 11.11 \\
        \bottomrule
    \end{tabular}

    \vspace{3pt}

    \begin{minipage}{0.98\textwidth}
        \footnotesize
        \textit{Note.} FT and NAD are omitted for Input-Aware because neither method yields a qualified representative state on any of the three datasets under the primary architectures.
    \end{minipage}
\end{table*}

\subsection{Repair Protocol}
\label{sec:repair_protocol}

For each dataset--architecture--attack--target configuration in the primary benchmark, we use a single fixed backdoored model as the pre-repair model $f_{\mathrm{pre}}$ for all repair methods and repair repetitions. Fixing the same pre-repair checkpoint removes variation due to pre-repair model construction and provides a common reference for preservation analysis within each configuration. Each repair method is evaluated over three repair repetitions, each using a trusted clean subset containing $5\%$ of the corresponding training set. Within each dataset and repair repetition, all methods use the same trusted subset, controlling repair-data variation across methods. The clean-label validation uses a separate repair-data configuration, as detailed in Appendix~\ref{app:clean_label_details}.

Because different repair methods operate through different intervention variables, they do not share a common notion of repair strength. We therefore define a method-specific set of candidate repaired states along each method's intervention dimension. For FT, NAD, and D3, candidates are collected at predefined stages of optimization; for FP and ANP, they correspond to different pruning settings; and for I-BAU, they correspond to different numbers of completed outer repair rounds. These candidate sets are predefined independently of preservation outcomes. A representative state is then selected from each candidate set using the common criterion defined in Section~\ref{sec:state_selection}. Detailed repair configurations and candidate-state settings are provided in Table~\ref{tab:app_repair_config} of Appendix~\ref{app:repair_details}.

\subsection{Representative Repair State Selection}
\label{sec:state_selection}

Within each repair repetition, let $S_r$ denote the predefined candidate set for repair method $r$, and let $f_r^{\mathrm{rep}}(s)$ denote the repaired model corresponding to candidate $s \in S_r$. Our objective is to identify a representative repair outcome using conventional evaluation criteria alone, without using any preservation-oriented information during selection.

We first require candidate states to satisfy a predefined attack-suppression criterion. Among the qualified candidates, we select the state with the highest Overall Clean Accuracy:
\begin{equation}
s^* =
\arg\max_{\substack{s\in S_r\\
\mathrm{ASR}(f_r^{\mathrm{rep}}(s))\le \tau}}
A_{\mathrm{overall}}(f_r^{\mathrm{rep}}(s)).
\label{eq:state_selection}
\end{equation}
The primary evaluation sets $\tau=0.10$, and the selection is performed independently within each repair repetition. This criterion first restricts the analysis to outcomes that achieve effective attack suppression and then favors the qualified state with the strongest aggregate clean utility.

Only after the representative state has been fixed do we evaluate benign-performance preservation relative to the corresponding pre-repair model. Preservation-oriented quantities do not participate in representative-state selection. For the selected state, we compute the class-wise performance changes $R_c$ and preservation losses $D_c$, and characterize localized preservation loss using the complete class-wise profile together with $D_{\max}$ and $D_{\mathrm{tail}}^{(\alpha)}$. The primary analysis uses $\alpha=0.8$. In this way, any preservation loss identified in the subsequent analysis is revealed only after a repair outcome has already been selected using conventional criteria.

If no candidate in a repair repetition satisfies the ASR criterion, that repetition has no qualified representative state. Condition-level results are averaged over the repetitions with qualified representative states. If none of the three repetitions qualifies, the corresponding condition is reported as NQ.

\section{Empirical Analysis}
\label{sec:empirical_analysis}

We empirically evaluate benign-performance preservation after backdoor repair and its relation to conventional aggregate clean evaluation. We first examine whether favorable aggregate clean outcomes can coexist with substantial localized preservation loss, and then analyze the class-wise structure of these losses. We further assess the robustness and generality of the observations across attack targets, model architectures, and class-weighting schemes, followed by additional validation under clean-label attacks.

Unless otherwise specified, the primary analysis uses target class $0$ and $D_{\mathrm{tail}}^{(.8)}$, corresponding to the upper 20\% of the class-wise preservation-loss distribution. All reported repair outcomes follow the representative-state selection protocol defined in Section~\ref{sec:state_selection}.

\begin{figure*}[t]
    \centering
    \includegraphics[width=\textwidth]{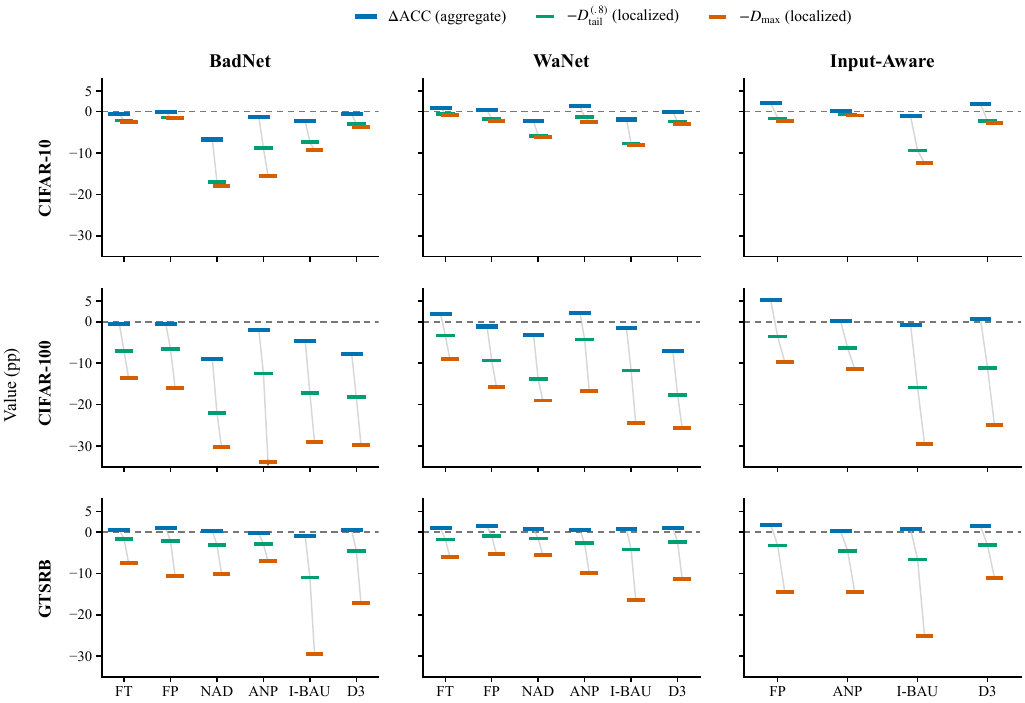}
    \caption{
Aggregate clean-performance change and localized preservation loss under the primary target-0 setting. Localized losses are plotted below zero. For CIFAR-100/BadNet/ANP, $D_{\max}=52.00$ pp exceeds the range.
}
\label{fig:aggregate_localized}
\end{figure*}

\subsection{Aggregate Utility Can Obscure Localized Preservation Loss}
\label{sec:aggregate_localized}

Table~\ref{tab:main_preservation_results} and Figure~\ref{fig:aggregate_localized} characterize the relationship between aggregate clean-performance change and localized preservation loss under the primary target-0 setting. Across the evaluated datasets, attacks, and repair methods, the two quantities frequently differ substantially in magnitude. In particular, modest changes in Overall Clean Accuracy can accompany considerably larger losses on the most affected classes. Under BadNet, for example, ANP changes Overall Clean Accuracy by -1.20 pp on CIFAR-10 while yielding $D_{\mathrm{tail}}^{(.8)}=8.78$ pp and $D_{\max}=15.53$ pp. On CIFAR-100, the same attack--repair combination produces a $\Delta$ACC of -1.91 pp, whereas the corresponding tail and worst-class losses reach 12.43 pp and 52.00 pp, respectively. Thus, localized degradation can be substantially attenuated when viewed only through the aggregate clean outcome.

This discrepancy is not limited to repair outcomes with decreased aggregate clean performance. In several conditions, Overall Clean Accuracy improves after repair while nontrivial preservation loss remains. Under Input-Aware on CIFAR-100, FP increases Overall Clean Accuracy by 5.17 pp, yet $D_{\mathrm{tail}}^{(.8)}$ and $D_{\max}$ remain 3.60 pp and 9.67 pp, respectively. On GTSRB, I-BAU similarly improves Overall Clean Accuracy by 0.69 pp under Input-Aware while incurring an upper-tail loss of 6.60 pp and a worst-class loss of 25.19 pp. Under WaNet on CIFAR-100, ANP yields a 2.20 pp aggregate improvement while its worst-class preservation loss reaches 16.67 pp. These outcomes are consistent with the aggregation effects analyzed in Section~\ref{sec:aggregate_preservation}: gains on some classes can offset losses on others, leaving substantial localized degradation weakly reflected in the aggregate result. The magnitude of the aggregate--localized discrepancy also varies across repair methods. As shown in Figure~\ref{fig:aggregate_localized}, repair outcomes with similar changes in Overall Clean Accuracy can exhibit markedly different upper-tail and worst-class preservation losses. Thus, aggregate clean utility does not uniquely characterize the preservation outcome of a repair.

\begin{figure*}[t]
    \centering
    \includegraphics[width=\textwidth]{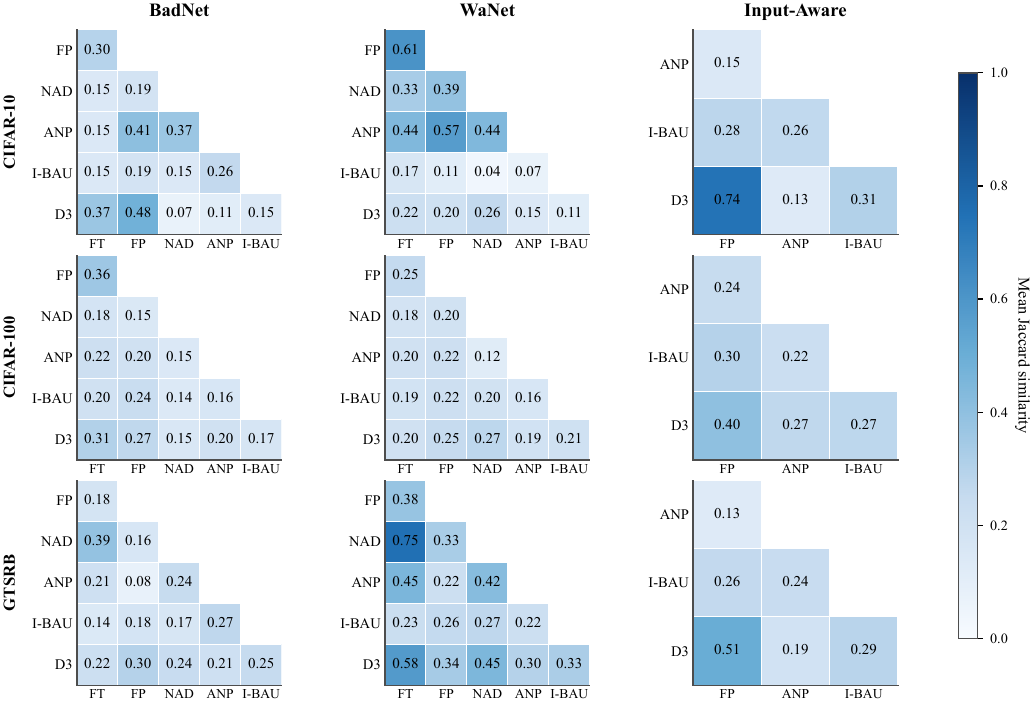}
    \caption{Cross-repair similarity of severely affected class sets $\mathcal{T}_{0.2}$ under the primary target-0 setting. Each cell reports the mean Jaccard similarity for a unique repair-method pair.}
    \label{fig:cross_repair_jaccard}
\end{figure*}

\newcommand{\qrep}[2]{#1\hspace{0.15em}/\hspace{0.15em}#2}

\begin{table*}[t]
\centering
\caption{
Attack-target and maximum non-target preservation loss, together with repetition-level target occurrence among the most affected classes, under the primary target-0 setting.
}
\label{tab:target_relation}

\setlength{\tabcolsep}{3.2pt}
\renewcommand{\arraystretch}{1.08}

\begin{tabular}{ll*{2}{rrrrrr}rrrr}
    \toprule
    \multirow{2}{*}{Dataset}
    & \multirow{2}{*}{Metric}
    & \multicolumn{6}{c}{BadNet}
    & \multicolumn{6}{c}{WaNet}
    & \multicolumn{4}{c}{Input-Aware} \\
    \cmidrule(lr){3-8}
    \cmidrule(lr){9-14}
    \cmidrule(lr){15-18}

    &
    & FT & FP & NAD & ANP & I-BAU & D3
    & FT & FP & NAD & ANP & I-BAU & D3
    & FP & ANP & I-BAU & D3 \\
    \midrule

    \multirow{4}{*}{CIFAR-10}
    & $D_t$ (pp)
    & 1.97 & 1.23 & 12.10 & 15.53 & 3.50 & 1.13
    & 0.35 & 2.03 & 6.20 & 2.47 & 2.27 & 1.10
    & 0.00 & 0.93 & 8.95 & 0.00 \\

    & $D_{\max}^{\neg t}$ (pp)
    & 2.07 & 1.47 & 17.17 & 2.03 & 9.23 & 3.77
    & 0.75 & 1.40 & 5.43 & 0.07 & 8.17 & 2.90
    & 2.17 & 0.37 & 7.60 & 2.63 \\

    \addlinespace[1.5pt]

    & Target-Worst
    & \qrep{1}{3} & \qrep{1}{3} & \qrep{1}{3}
    & \qrep{3}{3} & \qrep{0}{3} & \qrep{0}{3}
    & \qrep{0}{2} & \qrep{2}{3} & \qrep{2}{3}
    & \qrep{3}{3} & \qrep{0}{3} & \qrep{0}{3}
    & \qrep{0}{3} & \qrep{3}{3} & \qrep{1}{2} & \qrep{0}{3} \\

    & Target-Top20
    & \qrep{1}{3} & \qrep{2}{3} & \qrep{2}{3}
    & \qrep{3}{3} & \qrep{1}{3} & \qrep{0}{3}
    & \qrep{2}{2} & \qrep{3}{3} & \qrep{3}{3}
    & \qrep{3}{3} & \qrep{0}{3} & \qrep{1}{3}
    & \qrep{0}{3} & \qrep{3}{3} & \qrep{1}{2} & \qrep{0}{3} \\
    \midrule

    \multirow{4}{*}{CIFAR-100}
    & $D_t$ (pp)
    & 2.00 & 3.33 & 4.00 & 52.00 & 8.00 & 2.33
    & 2.67 & 2.67 & 0.00 & 16.67 & 8.50 & 5.67
    & 0.00 & 1.00 & 2.50 & 1.00 \\

    & $D_{\max}^{\neg t}$ (pp)
    & 13.67 & 16.00 & 30.33 & 24.33 & 29.00 & 29.67
    & 9.00 & 15.67 & 19.00 & 10.33 & 24.50 & 25.67
    & 9.67 & 11.33 & 29.50 & 25.00 \\

    \addlinespace[1.5pt]

    & Target-Worst
    & \qrep{0}{3} & \qrep{0}{3} & \qrep{0}{3}
    & \qrep{3}{3} & \qrep{0}{3} & \qrep{0}{3}
    & \qrep{0}{3} & \qrep{0}{3} & \qrep{0}{1}
    & \qrep{3}{3} & \qrep{0}{2} & \qrep{0}{3}
    & \qrep{0}{3} & \qrep{0}{3} & \qrep{0}{2} & \qrep{0}{3} \\

    & Target-Top20
    & \qrep{1}{3} & \qrep{1}{3} & \qrep{0}{3}
    & \qrep{3}{3} & \qrep{1}{3} & \qrep{0}{3}
    & \qrep{3}{3} & \qrep{0}{3} & \qrep{0}{1}
    & \qrep{3}{3} & \qrep{1}{2} & \qrep{0}{3}
    & \qrep{0}{3} & \qrep{0}{3} & \qrep{0}{2} & \qrep{0}{3} \\
    \midrule

    \multirow{4}{*}{GTSRB}
    & $D_t$ (pp)
    & 0.56 & 0.00 & 2.78 & 0.00 & 1.11 & 0.00
    & 0.00 & 0.00 & 0.00 & 0.00 & 0.00 & 0.00
    & 0.00 & 0.00 & 5.56 & 0.00 \\

    & $D_{\max}^{\neg t}$ (pp)
    & 7.52 & 10.74 & 9.63 & 7.12 & 29.44 & 17.22
    & 6.00 & 5.33 & 5.56 & 10.00 & 16.39 & 11.33
    & 14.44 & 14.44 & 25.19 & 11.11 \\

    \addlinespace[1.5pt]

    & Target-Worst
    & \qrep{0}{3} & \qrep{0}{3} & \qrep{1}{3}
    & \qrep{0}{3} & \qrep{0}{3} & \qrep{0}{3}
    & \qrep{0}{1} & \qrep{0}{3} & \qrep{0}{1}
    & \qrep{0}{3} & \qrep{0}{3} & \qrep{0}{3}
    & \qrep{0}{3} & \qrep{0}{3} & \qrep{0}{3} & \qrep{0}{3} \\

    & Target-Top20
    & \qrep{1}{3} & \qrep{0}{3} & \qrep{2}{3}
    & \qrep{0}{3} & \qrep{0}{3} & \qrep{0}{3}
    & \qrep{0}{1} & \qrep{0}{3} & \qrep{0}{1}
    & \qrep{0}{3} & \qrep{0}{3} & \qrep{0}{3}
    & \qrep{0}{3} & \qrep{0}{3} & \qrep{1}{3} & \qrep{0}{3} \\
    \bottomrule
\end{tabular}

\vspace{3pt}

\begin{minipage}{0.98\textwidth}
\footnotesize
\textit{Note.}
For Target-Worst and Target-Top20, $k/n$ indicates that the attack-target class is, respectively, among the worst-loss classes or within the severely affected class set $T_{0.2}$ in $k$ of the $n$ qualified repair repetitions. 
\end{minipage}

\end{table*}

\begin{figure*}[t]
    \centering
    \includegraphics[width=\textwidth]{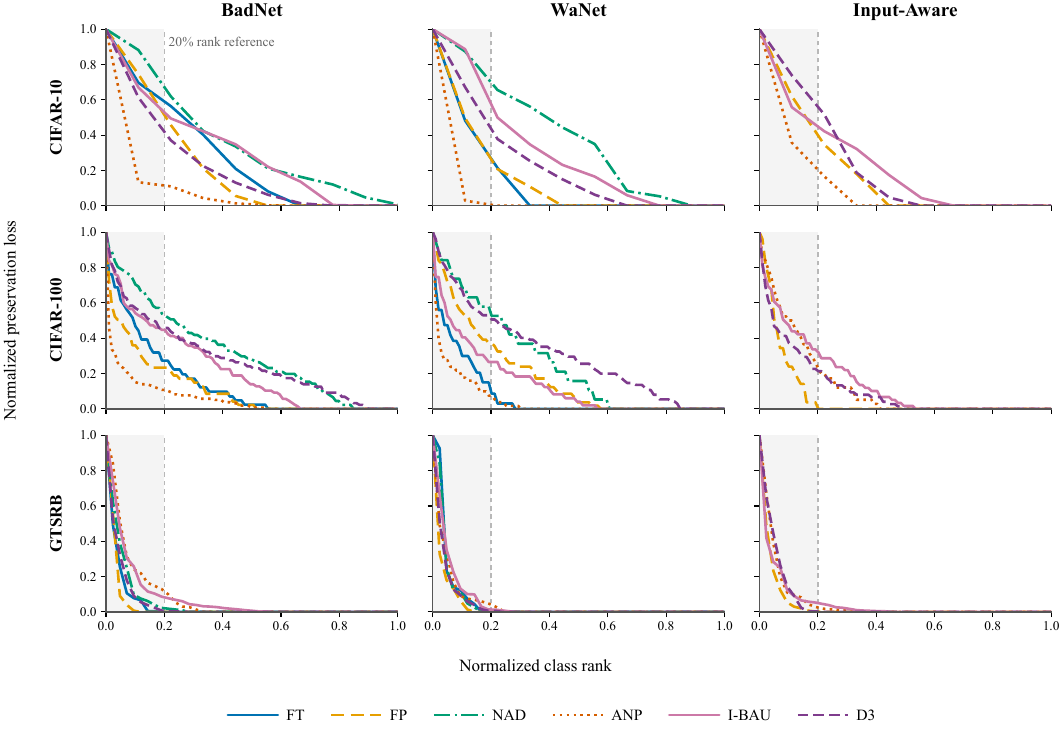}
   \caption{Ranked class-wise preservation-loss profiles under the primary target-0 setting. Each curve averages repetition-wise profiles obtained by sorting class-wise losses and normalizing them by the corresponding worst-class loss. The vertical reference marks the most affected 20\% of the class space.}
    \label{fig:ranked_profiles}
\end{figure*}

\subsection{Class-Wise Structure of Preservation Loss}
\label{sec:classwise_structure}

Having established the aggregate--localized discrepancy, we next examine the class-wise structure of preservation loss. We consider three questions: whether the most severely affected class is systematically the attack target, whether the affected classes are consistent across repair methods, and whether preservation loss is concentrated in a small number of classes or distributed more broadly across the label space. For the following class-set analyses, we define $T_{0.2}$ as the positive-loss classes within the most affected 20\% of the class space.

\paragraph{Attack Target and Worst-Class Preservation Loss}

We first examine whether the attack target systematically coincides with the most severely affected class. Table~\ref{tab:target_relation} reports four complementary quantities. $D_t$ denotes the preservation loss of the attack-target class. Target-Worst reports $k/n$, where $n$ is the number of qualified repair repetitions and $k$ is the number of repetitions in which the target attains the largest preservation loss. Target-Top20 analogously reports the number of qualified repetitions in which the target belongs to the severely affected class set $T_{0.2}$. Finally, $D_{\max}^{\neg t}$ denotes the largest preservation loss among non-target classes. Across the 48 reported dataset--attack--repair conditions, Target-Worst records no such occurrence in 36 cases, showing that the attack target is not systematically the worst-loss class across the benchmark.

Target-centered degradation nevertheless appears clearly under particular repair conditions. ANP provides the most prominent examples: on CIFAR-100, the target is among the worst-loss classes in every qualified repetition under both BadNet and WaNet, with corresponding $D_t$ values of 52.00 pp and 16.67 pp, respectively. Similar target-centered behavior is also observed in several CIFAR-10 conditions. These results confirm that degradation of the attack-target class can constitute an important class-wise effect of repair.

However, substantial preservation loss can also arise when the target is not among the most affected classes. Under CIFAR-100/Input-Aware, both Target-Worst and Target-Top20 record zero occurrences for every reported repair method, while $D_{\max}^{\neg t}$ ranges from 9.67 pp to 29.50 pp. The contrast is even clearer on GTSRB/WaNet: the target incurs no preservation loss for any repair method, yet the largest non-target losses range from 5.33 pp to 16.39 pp. Thus, target-class degradation is an important manifestation of localized preservation loss under some repair conditions, but it does not generally explain the most severe class-wise losses; substantial preservation loss can also occur among non-target classes.

\paragraph{Affected-Class Agreement Across Repair Methods}
We next examine whether different repair methods consistently affect the same classes. For each pair of repair methods, we compute the Jaccard similarity between their $T_{0.2}$ sets across qualified repair repetitions and report the mean pairwise similarity in Figure~\ref{fig:cross_repair_jaccard}.

Overall, cross-repair agreement is generally limited rather than uniformly high. Across all off-diagonal repair-pair comparisons, the median Jaccard similarity is 0.22, with values ranging from 0.04 to 0.75. The pattern is particularly clear on CIFAR-100, where all off-diagonal similarities remain at or below 0.40 across the three attacks. At the same time, substantial agreement can occur for particular repair pairs. Under CIFAR-10/Input-Aware, for example, FP and D3 reach 0.74, whereas ANP and D3 reach only 0.13. Under CIFAR-10/WaNet, FT and FP reach 0.61, while NAD and I-BAU overlap by only 0.04. GTSRB/WaNet exhibits a similar contrast, with FT and NAD reaching 0.75 while several other repair pairs show substantially lower agreement.

These results show that affected-class agreement is repair-pair specific rather than shared uniformly across repair methods. Even within the same dataset--attack condition, some repair pairs exhibit substantial overlap while others affect markedly different class sets. Thus, the severely affected classes do not form a common subset shared across the evaluated repair methods.

\paragraph{Concentration of Preservation Loss Across Classes}
Finally, we examine whether preservation loss is concentrated in a small number of classes or extends across a broader portion of the label space. For each qualified repair repetition, we sort the class-wise preservation losses in descending order and normalize the loss magnitude and class rank as $\widetilde{D}_{(i)}=D_{(i)}/D_{\max}$ and $\rho_i=(i-1)/(C-1)$, respectively. A rapidly decaying ranked profile indicates that preservation loss is concentrated in a small number of classes, whereas a slower decay indicates that the loss extends across a broader portion of the class space.

Figure~\ref{fig:ranked_profiles} shows that the concentration of preservation loss varies substantially across experimental conditions. The contrast is particularly visible across datasets. On GTSRB, most profiles decay sharply within the early portion of the class ranking, indicating that the loss is largely concentrated in a relatively small subset of classes. In contrast, several CIFAR-10 and CIFAR-100 conditions exhibit much longer profiles, with non-negligible normalized loss extending well beyond the 20\% rank reference. The profile shape also varies within the same dataset--attack condition. For example, under CIFAR-10/BadNet, ANP exhibits a sharply concentrated profile, whereas NAD distributes substantial relative loss across a much broader range of classes. Similar cross-repair differences appear under other attack settings.

Viewed together with the absolute losses reported in Table~\ref{tab:main_preservation_results}, the ranked profiles show that loss severity and breadth need not vary together. Some repair outcomes exhibit a large preservation loss concentrated in only a few classes, whereas others show more moderate losses extending across a broader portion of the label space. Thus, localized preservation loss differs not only in magnitude but also in how broadly it is distributed across classes.

\subsection{Robustness and Generalization}
\label{sec:generalization}

We next examine whether the aggregate--localized discrepancy depends on specific choices in the primary evaluation. We consider three potential sources of dependence: the attack target, the model architecture, and the class-weighting scheme used to summarize clean performance.

\begin{table}[t]
\centering
\caption{
Aggregate clean-performance change and tail preservation loss across
attack targets on CIFAR-100.
}
\label{tab:target_generalization_c100}

\renewcommand{\arraystretch}{1.05}
\setlength{\tabcolsep}{3.5pt}
\footnotesize

\begin{tabular}{
ll
>{\columncolor{convbg}}c
>{\columncolor{presbg}[\tabcolsep][2.75pt]}c
@{\hspace{7pt}}
>{\columncolor{convbg}[2.75pt][\tabcolsep]}c
>{\columncolor{presbg}[\tabcolsep][2.75pt]}c
@{\hspace{7pt}}
>{\columncolor{convbg}[2.75pt][\tabcolsep]}c
>{\columncolor{presbg}}c
}
\toprule

& &
\multicolumn{2}{c}{$t=0$} &
\multicolumn{2}{c}{$t=3$} &
\multicolumn{2}{c}{$t=6$} \\
\cmidrule(lr){3-4}
\cmidrule(lr){5-6}
\cmidrule(lr){7-8}

Attack & Repair
& $\Delta$ACC
& $D_{\mathrm{tail}}^{(.8)}$
& $\Delta$ACC
& $D_{\mathrm{tail}}^{(.8)}$
& $\Delta$ACC
& $D_{\mathrm{tail}}^{(.8)}$ \\
\midrule

\multirow{6}{*}{BadNet}
& FT
& -0.58 & 7.05
& -0.80 & 6.85
& -0.52 & 6.48 \\

& FP
& -0.55 & 6.53
& -0.55 & 6.57
& -0.38 & 6.20 \\

& NAD
& -8.98 & 21.98
& -11.91 & 25.17
& -10.32 & 25.43 \\

& ANP
& -1.91 & 12.43
& -2.32 & 12.58
& -2.90 & 16.48 \\

& I-BAU
& -4.60 & 17.25
& -5.37 & 19.58
& -5.36 & 20.83 \\

& D3
& -7.72 & 18.22
& -8.16 & 17.53
& -8.07 & 18.17 \\

\midrule

\multirow{6}{*}{WaNet}
& FT
& 1.98 & 3.27
& 3.13 & 3.20
& 1.88 & 3.75 \\

& FP
& -1.14 & 9.30
& -0.83 & 10.88
& -1.05 & 9.82 \\

& NAD
& -3.14 & 13.85
& -4.67 & 17.38
& -3.57 & 15.62 \\

& ANP
& 2.20 & 4.32
& 1.39 & 8.30
& 1.71 & 7.32 \\

& I-BAU
& -1.46 & 11.73
& -0.17 & 10.75
& -1.63 & 12.45 \\

& D3
& -7.07 & 17.63
& -6.33 & 17.70
& -7.23 & 18.28 \\

\midrule

\multirow{4}{*}{\shortstack[c]{Input-\\Aware}}
& FP
& 5.17 & 3.60
& 4.09 & 5.00
& 5.09 & 4.88 \\

& ANP
& 0.22 & 6.35
& -0.02 & 2.13
& 0.11 & 2.65 \\

& I-BAU
& -0.69 & 15.88
& 0.32 & 17.00
& -0.86 & 22.05 \\

& D3
& 0.72 & 11.13
& 0.11 & 11.32
& 0.65 & 12.90 \\

\bottomrule
\end{tabular}

\vspace{2pt}

\begin{minipage}{0.97\columnwidth}
\footnotesize
\textit{Note.}
All values are reported in percentage points (pp).
\end{minipage}

\end{table}

\paragraph{Attack-Target Generalization}
We next examine whether the aggregate--localized discrepancy observed in the primary experiments is specific to the choice of attack target. Table~\ref{tab:target_generalization_c100} reports results on CIFAR-100 for target classes $t\in\{0,3,6\}$ under the same experimental protocol.

The central pattern persists across all three target choices. For each target, multiple repair conditions exhibit non-decreasing Overall Clean Accuracy while retaining positive upper-tail preservation loss. Under WaNet/ANP, for example, $\Delta$ACC remains positive at 2.20, 1.39, and 1.71 pp for $t=0$, $3$, and $6$, respectively, while $D_{\mathrm{tail}}^{(.8)}$ reaches 4.32, 8.30, and 7.32 pp. An even stronger separation appears for Input-Aware/D3: $\Delta$ACC remains between 0.11 and 0.72 pp across the three targets, whereas $D_{\mathrm{tail}}^{(.8)}$ remains between 11.13 and 12.90 pp. Input-Aware/FP likewise improves Overall Clean Accuracy by 4.09--5.17 pp across all three targets while retaining tail losses of 3.60--5.00 pp.

The attack target nevertheless affects the quantitative preservation outcome. For example, $D_{\mathrm{tail}}^{(.8)}$ for BadNet/ANP varies from 12.43 pp at $t=0$ to 16.48 pp at $t=6$, while Input-Aware/I-BAU varies from 15.88 pp to 22.05 pp. Thus, the magnitude of localized preservation loss is target dependent, but the aggregate--localized discrepancy observed in the primary setting is not specific to target class $0$.

\paragraph{Architecture Generalization}
We next examine whether the aggregate--localized discrepancy observed in the primary experiments is specific to the ResNet18 architecture. Table~\ref{tab:architecture_generalization_c100} compares ResNet18, PreActResNet18, and VGG19-BN on CIFAR-100 under the same representative-state selection protocol. NQ denotes a setting in which no candidate repair state satisfies the ASR qualification criterion.

The central pattern is reproduced across all three architectures. Each architecture contains multiple repair outcomes for which Overall Clean Accuracy is non-decreasing or changes only slightly, while a nontrivial upper-tail preservation loss remains. Under BadNet/FP, for example, $\Delta$ACC is -0.55, -0.08, and 0.35 pp for ResNet18, PreActResNet18, and VGG19-BN, respectively, whereas $D_{\mathrm{tail}}^{(.8)}$ remains 6.53, 5.12, and 5.35 pp. Under Input-Aware/ANP, the aggregate changes are similarly small at 0.22, 0.09, and 0.14 pp, while the corresponding tail losses are 6.35, 1.55, and 3.72 pp. The aggregate--localized discrepancy therefore remains observable beyond the primary ResNet18 architecture.

At the same time, architecture can substantially change the quantitative repair outcome. Under WaNet/ANP, for example, $D_{\mathrm{tail}}^{(.8)}$ varies from 2.63 pp on VGG19-BN to 15.42 pp on PreActResNet18, accompanied by changes in the sign and magnitude of $\Delta$ACC. Several attack--repair combinations also yield NQ on one architecture while producing qualified repair states on another. Thus, architecture affects the severity and qualification of individual repair outcomes, but does not eliminate the aggregate--localized discrepancy observed in the primary evaluation.

\paragraph{Robustness to Class Weighting}
Overall Clean Accuracy weights classes according to their empirical test-set frequencies, whereas our preservation summaries treat classes uniformly. On class-imbalanced datasets, this difference in weighting could contribute to the observed aggregate--localized discrepancy. We therefore examine whether the discrepancy persists when aggregate clean performance is also computed with uniform class weighting.

Specifically, we use Macro Clean Accuracy and measure its repair-induced change as
\begin{equation}
\begin{aligned}
A_{\mathrm{macro}}(f)
&=
\frac{1}{C}\sum_{c=1}^{C} A_c(f), \\
\Delta A_{\mathrm{macro}}
&=
A_{\mathrm{macro}}(f^{\mathrm{rep}})
-
A_{\mathrm{macro}}(f^{\mathrm{pre}}).
\end{aligned}
\label{eq:macro_clean_accuracy}
\end{equation}
The representative repair states remain unchanged from the primary evaluation and are not reselected using Macro Clean Accuracy. For CIFAR-10 and CIFAR-100, the test sets are class balanced, so Overall Clean Accuracy and Macro Clean Accuracy coincide by construction. The aggregate--localized discrepancy observed on these datasets therefore already arises without any class-frequency weighting mismatch.

We further evaluate this factor on the class-imbalanced GTSRB dataset in Table~\ref{tab:macro_acc_control}. Replacing sample-weighted Overall Clean Accuracy with class-uniform Macro Clean Accuracy changes the aggregate clean-performance change only modestly. Across the 16 qualified attack--repair conditions, the mean absolute difference between $\Delta$ACC and $\Delta A_{\mathrm{macro}}$ is 0.46 pp, with a maximum difference of 0.88 pp. Only two conditions change sign, and both remain close to zero: BadNet/ANP changes from -0.33 pp to 0.24 pp, while Input-Aware/ANP changes from 0.20 pp to -0.27 pp.

More importantly, the aggregate--localized discrepancy persists under uniform class weighting. Fourteen of the sixteen qualified conditions exhibit non-decreasing Macro Clean Accuracy, yet all retain positive upper-tail preservation loss. Under BadNet/D3, for example, $\Delta A_{\mathrm{macro}}$ is 0.86 pp while $D_{\mathrm{tail}}^{(.8)}$ reaches 4.54 pp. Under WaNet/I-BAU, the corresponding values are 1.31 pp and 4.25 pp, and under Input-Aware/I-BAU, Macro Clean Accuracy still increases by 0.14 pp while the upper-tail preservation loss reaches 6.60 pp. Thus, the observed discrepancy cannot be explained by class-frequency weighting alone: localized preservation loss can remain weakly reflected in an aggregate clean-performance statistic even when both are evaluated with uniform class weighting.

\begin{table}[t]
\centering
\caption{
Aggregate clean-performance change and tail preservation loss across
model architectures on CIFAR-100.
}
\label{tab:architecture_generalization_c100}

\renewcommand{\arraystretch}{1.05}
\setlength{\tabcolsep}{3.5pt}
\footnotesize

\begin{tabular}{
ll
>{\columncolor{convbg}}c
>{\columncolor{presbg}[\tabcolsep][2.75pt]}c
@{\hspace{7pt}}
>{\columncolor{convbg}[2.75pt][\tabcolsep]}c
>{\columncolor{presbg}[\tabcolsep][2.75pt]}c
@{\hspace{7pt}}
>{\columncolor{convbg}[2.75pt][\tabcolsep]}c
>{\columncolor{presbg}}c
}
\toprule

& &
\multicolumn{2}{c}{ResNet18} &
\multicolumn{2}{c}{PreActResNet18} &
\multicolumn{2}{c}{VGG19-BN} \\
\cmidrule(lr){3-4}
\cmidrule(lr){5-6}
\cmidrule(lr){7-8}

Attack & Repair
& $\Delta$ACC
& $D_{\mathrm{tail}}^{(.8)}$
& $\Delta$ACC
& $D_{\mathrm{tail}}^{(.8)}$
& $\Delta$ACC
& $D_{\mathrm{tail}}^{(.8)}$ \\
\midrule

\multirow{6}{*}{BadNet}
& FT
& -0.58 & 7.05
& -0.25 & 5.52
& -1.27 & 9.12 \\

& FP
& -0.55 & 6.53
& -0.08 & 5.12
& 0.35 & 5.35 \\

& NAD
& -8.98 & 21.98
& -0.49 & 6.15
& -2.42 & 12.08 \\

& ANP
& -1.91 & 12.43
& -0.45 & 8.20
& -0.48 & 11.43 \\

& I-BAU
& -4.60 & 17.25
& -3.91 & 16.93
& -2.62 & 15.48 \\

& D3
& -7.72 & 18.22
& -6.64 & 15.00
& 0.92 & 4.63 \\

\midrule

\multirow{6}{*}{WaNet}
& FT
& 1.98 & 3.27
& 1.43 & 3.48
& -2.86 & 12.05 \\

& FP
& -1.14 & 9.30
& -1.12 & 8.20
& -1.48 & 8.98 \\

& NAD
& -3.14 & 13.85
& 1.42 & 3.77
& -2.94 & 14.90 \\

& ANP
& 2.20 & 4.32
& -2.77 & 15.42
& 1.64 & 2.63 \\

& I-BAU
& -1.46 & 11.73
& \multicolumn{2}{c}{\textit{NQ}}
& -4.02 & 15.57 \\

& D3
& -7.07 & 17.63
& -7.56 & 17.00
& -1.79 & 10.38 \\

\midrule

\multirow{6}{*}{\shortstack[c]{Input-\\Aware}}
& FT
& \multicolumn{2}{c}{\textit{NQ}}
& \multicolumn{2}{c}{\textit{NQ}}
& 4.59 & 7.65 \\

& FP
& 5.17 & 3.60
& -0.40 & 12.25
& 4.88 & 6.22 \\

& NAD
& \multicolumn{2}{c}{\textit{NQ}}
& \multicolumn{2}{c}{\textit{NQ}}
& 4.40 & 7.15 \\

& ANP
& 0.22 & 6.35
& 0.09 & 1.55
& 0.14 & 3.72 \\

& I-BAU
& -0.69 & 15.88
& 1.65 & 8.95
& \multicolumn{2}{c}{\textit{NQ}} \\

& D3
& 0.72 & 11.13
& -0.10 & 11.20
& 4.64 & 7.00 \\

\bottomrule
\end{tabular}

\vspace{2pt}

\begin{minipage}{0.97\columnwidth}
\footnotesize
\textit{Note.}
All values are reported in percentage points (pp).
\textit{NQ} indicates that no qualified representative state is obtained.
\end{minipage}

\end{table}

\subsection{Additional Validation under Clean-Label Attacks}
\label{sec:clean_label_validation}
The primary experiments consider BadNet, WaNet, and Input-Aware, but do not explicitly cover clean-label poisoning. We therefore conduct complementary validation on LC and SIG using CIFAR-10 and include FST and RNP as additional repair methods. Following the corresponding experimental protocols, the repair procedures use a trusted clean subset containing 2\% of the training set. Representative repair states are selected using the same criterion as in Section~\ref{sec:state_selection}.

Table~\ref{tab:clean_label_validation} shows that the aggregate--localized discrepancy remains observable across all four attack--repair conditions. All reported representative repair outcomes satisfy the ASR qualification criterion, with condition-level ASR ranging from 0.36\% to 9.25\%. The largest discrepancy in magnitude appears under LC/RNP: Overall Clean Accuracy decreases by 2.05 pp, while $D_{\mathrm{tail}}^{(.8)}$ and $D_{\max}$ reach 11.35 pp and 13.80 pp, respectively. A complementary case appears under SIG/RNP, where Overall Clean Accuracy is nearly unchanged, with $\Delta$ACC of only -0.01 pp, yet the repaired model still exhibits a tail preservation loss of 1.00 pp and a worst-class loss of 1.47 pp. FST exhibits the same qualitative pattern under both attacks, with localized preservation losses exceeding the corresponding aggregate clean-performance changes.

\begin{table}[t]
\centering
\caption{
Aggregate clean-performance change under sample-weighted and
class-uniform evaluation, together with tail preservation loss,
on GTSRB.
}
\label{tab:macro_acc_control}

\renewcommand{\arraystretch}{1.05}
\setlength{\tabcolsep}{7pt}
\small

\begin{tabular}{
ll
>{\columncolor{convbg}}c
>{\columncolor{convbg}[\tabcolsep][6.25pt]}c
@{\hspace{14pt}}
>{\columncolor{presbg}[6.25pt][\tabcolsep]}c
}
\toprule

Attack & Repair
& $\Delta$ACC
& $\Delta A_{\mathrm{macro}}$
& $D_{\mathrm{tail}}^{(.8)}$ \\
\midrule

\multirow{6}{*}{BadNet}
& FT
& 0.47  & 0.76  & 1.64 \\

& FP
& 0.94  & 1.52  & 2.18 \\

& NAD
& 0.28  & 0.42  & 3.08 \\

& ANP
& -0.33 & 0.24  & 2.94 \\

& I-BAU
& -1.01 & -1.79 & 10.95 \\

& D3
& 0.44  & 0.86  & 4.54 \\
\midrule

\multirow{6}{*}{WaNet}
& FT
& 0.87 & 1.35 & 1.77 \\

& FP
& 1.34 & 1.87 & 0.94 \\

& NAD
& 0.78 & 1.15 & 1.55 \\

& ANP
& 0.39 & 1.27 & 2.72 \\

& I-BAU
& 0.74 & 1.31 & 4.25 \\

& D3
& 0.97 & 1.37 & 2.39 \\
\midrule

\multirow{4}{*}{\shortstack[c]{Input-\\Aware}}
& FP
& 1.75 & 1.89  & 3.28 \\

& ANP
& 0.20 & -0.27 & 4.56 \\

& I-BAU
& 0.69 & 0.14  & 6.60 \\

& D3
& 1.42 & 1.50  & 3.16 \\

\bottomrule
\end{tabular}

\vspace{2pt}

\begin{minipage}{0.97\columnwidth}
\small
\textit{Note.}
All values are reported in percentage points (pp).
\end{minipage}

\end{table}

\begin{table}[t]
\centering
\caption{Additional validation under clean-label attacks on CIFAR-10.}
\label{tab:clean_label_validation}

\setlength{\tabcolsep}{7.0pt}
\renewcommand{\arraystretch}{1.08}
\small

\begin{tabular}{@{}lrrrr@{}}
\toprule
&
\multicolumn{2}{c}{LC} &
\multicolumn{2}{c}{SIG} \\
\cmidrule(lr){2-3}
\cmidrule(lr){4-5}

Metric
& FST & RNP
& FST & RNP \\
\midrule

\rowcolor{convbg}
\cellcolor{white}
ASR (\%)
& 0.36 & 9.25
& 1.70 & 1.52 \\

\rowcolor{convbg}
\cellcolor{white}
$\Delta$ACC (pp)
& -0.40 & -2.05
& -0.78 & -0.01 \\

\addlinespace[1.5pt]

\rowcolor{presbg}
\cellcolor{white}
$D_{\mathrm{tail}}^{(.8)}$ (pp)
& 1.77 & 11.35
& 2.47 & 1.00 \\

\rowcolor{presbg}
\cellcolor{white}
$D_{\max}$ (pp)
& 2.03 & 13.80
& 2.77 & 1.47 \\

\bottomrule
\end{tabular}

\end{table}

These results provide complementary evidence that the aggregate--localized discrepancy observed in the primary experiments also arises under clean-label poisoning. Across both LC and SIG, localized preservation loss remains observable even when aggregate clean performance changes only modestly or is nearly unchanged.

\section{Conclusion}
\label{sec:conclusion}

This work revisits how benign-performance preservation is evaluated in backdoor repair. While Attack Success Rate and Overall Clean Accuracy remain useful for characterizing attack suppression and aggregate clean utility, respectively, aggregate clean performance does not fully characterize whether pre-repair class-wise benign performance is preserved. We therefore distinguish aggregate clean utility from benign-performance preservation and operationalize the latter through class-wise preservation loss, together with worst-class and upper-tail summaries. This preservation-oriented view complements conventional evaluation without collapsing attack suppression, aggregate utility, and localized preservation into a single score.

Our empirical analysis shows that repair outcomes selected under conventional criteria can still exhibit substantial localized preservation loss. This loss is also not confined to a single class-wise pattern: the most severely affected class is not systematically the attack target, and the breadth and location of degradation vary across repair outcomes. These findings show that ASR and aggregate clean accuracy alone can leave important changes in previously available benign performance uncharacterized. Reporting localized preservation information alongside conventional security and utility metrics therefore provides a more complete account of backdoor repair outcomes. Our study focuses on class-wise performance preservation in image classification. Extending this perspective to finer-grained subpopulations, other prediction tasks, and broader model families offers a natural direction for future work.

 

\bibliographystyle{IEEEtran}
\bibliography{ref_clean}

\appendix

\subsection{Experimental Protocol}
\label{app:experimental_protocol}

Unless otherwise specified, the following details apply to the primary benchmark. For each dataset--architecture--attack--target configuration, the pre-repair model is trained with seed 1 and the resulting checkpoint is kept fixed across all repair methods and repair repetitions. The clean accuracy and ASR of the fixed pre-repair models used in the primary target-0 benchmark are reported in Table~\ref{tab:app_pre_repair_performance}. Each repair method is evaluated over three repetitions using seeds 1, 2, and 3. In each repetition, a trusted clean subset containing 5\% of the corresponding training set is sampled without replacement and without explicit class balancing. Within the same dataset and repetition, all repair methods use the same trusted subset.

ASR is computed over test samples whose ground-truth labels differ from the designated attack target. Representative-state selection follows Section~\ref{sec:state_selection} and uses the exact, unrounded ASR and Overall Clean Accuracy values. Exact ties are resolved according to the predefined candidate order. Method-specific candidate-state construction is detailed in Appendix~\ref{app:repair_details}.

\subsection{Pre-Repair Model Construction and Attack Instantiation}
\label{app:pre_repair_attack}

\begin{table}[t]
\centering
\caption{
Performance of the fixed pre-repair models used in the primary
target-0 benchmark (seed 1).
}
\label{tab:app_pre_repair_performance}

\small
\renewcommand{\arraystretch}{1.05}
\setlength{\tabcolsep}{7pt}

\begin{tabular}{clcc}
\toprule
Dataset & Attack & Clean ACC (\%) & ASR (\%) \\
\midrule

\multirow{3}{*}{CIFAR-10}
& BadNet      & 90.96 & 94.51 \\
& WaNet       & 91.65 & 99.53 \\
& Input-Aware & 90.71 & 95.06 \\

\midrule

\multirow{3}{*}{CIFAR-100}
& BadNet      & 66.39 & 88.93 \\
& WaNet       & 69.72 & 99.40 \\
& Input-Aware & 63.57 & 86.25 \\

\midrule

\multirow{3}{*}{GTSRB}
& BadNet      & 97.51 & 94.81 \\
& WaNet       & 97.22 & 99.71 \\
& Input-Aware & 96.55 & 92.35 \\

\bottomrule
\end{tabular}
\end{table}

We use the standard training and test partitions of CIFAR-10, CIFAR-100, and GTSRB, with all inputs processed at a resolution of $32\times32$. The primary benchmark uses ResNet18-CIFAR on CIFAR-10 and CIFAR-100 and PreActResNet18 on GTSRB. For the architecture-generalization analysis on CIFAR-100, we additionally consider PreActResNet18 and VGG19-BN.

The primary benchmark considers BadNet, WaNet, and Input-Aware in the all-to-one setting. Class $0$ is used as the primary attack target. For the target-generalization analysis on CIFAR-100, we additionally consider target classes $3$ and $6$.

For BadNet and WaNet, the pre-repair models are trained using SGD with an initial learning rate of $0.01$, momentum $0.9$, weight decay $5\times10^{-4}$, and a batch size of 128. The CIFAR-10 and CIFAR-100 models are trained for 100 epochs with cosine annealing, whereas the GTSRB models are trained for 50 epochs. The additional PreActResNet18 and VGG19-BN models on CIFAR-100 follow the same training configuration as the primary CIFAR-100 models. Input-Aware instead follows its attack-specific joint-training procedure described below.

\paragraph{BadNet}
We use a fixed $3\times3$ white-square trigger placed in the bottom-right corner of the image. The poisoning ratio is $10\%$, and poisoned samples are relabeled to the designated target class.

\paragraph{WaNet}
WaNet generates triggered samples through smooth spatial warping. We use a poisoning ratio of $0.1$ and a cross ratio of $2.0$, with twice as many cross samples as backdoor samples. The warping strength is set to $s=0.5$, with kernel size $k=4$ and grid-rescaling factor $1.0$. Backdoor and cross samples are generated online during training, and the corresponding warping grid is reused during evaluation.

\paragraph{Input-Aware}
Input-Aware generates sample-dependent trigger patterns and masks using dedicated generator networks. The mask generator is pretrained for 25 epochs using Adam with a learning rate of $0.01$ and is then frozen. The classifier and trigger generator are subsequently trained jointly for 75 epochs on CIFAR-10 and CIFAR-100 and for 25 epochs on GTSRB, using a batch size of 128. Backdoor and cross samples are each generated using a nominal ratio of $0.1$. The classifier is optimized using SGD with a learning rate of $0.01$, while the trigger generator is optimized using Adam with a learning rate of $0.01$.

\subsection{Repair Instantiation and Candidate-State Construction}
\label{app:repair_details}

The primary benchmark evaluates FT, FP, NAD, ANP, I-BAU, and D3. All repair methods use a batch size of 256. Table~\ref{tab:app_repair_config} summarizes the main repair configurations and predefined candidate states considered for each method. These candidate sets are fixed independently of the preservation-oriented quantities introduced in Section~\ref{sec:preservation}. The following paragraphs provide the main settings that determine the repair trajectory, candidate-state construction, and architecture-specific instantiation.

\begin{table*}[t]
\centering
\caption{Repair configurations and predefined candidate states used in the primary benchmark.}
\label{tab:app_repair_config}
\small
\begin{tabular}{llll}
\hline
Method & Repair configuration & Candidate variable & Candidate states \\
\hline

FT
& Full-model fine-tuning for up to 100 epochs
& Fine-tuning epoch
& $\{1,2,5,10,20,40,60,80,100\}$ \\

FP
& Activation-based pruning followed by 100-epoch fine-tuning
& Pruning fraction
& $\{0.1,0.3,0.5,0.7,0.8,0.9,0.95\}$ \\

NAD
& 10-epoch teacher fine-tuning and up to 20 student epochs
& Student epoch
& $\{1,2,4,6,8,10,20\}$ \\

ANP
& 2,000 adversarial mask-optimization iterations
& Pruning threshold
& $\{0,0.1,0.2,0.3,0.4,0.5,0.6,0.7\}$ \\

I-BAU
& Iterative adversarial unlearning for 10 outer rounds
& Completed outer round
& $\{1,2,3,4,5,8,10\}$ \\

D3
& Distance-driven optimization for up to 100 epochs
& Optimization epoch
& $\{1,2,5,10,20,40,60,80,100\}$ \\

\hline
\end{tabular}
\end{table*}

FT, NAD, and D3 obtain candidate states from a single continuous optimization trajectory. For FP, one activation-based channel ranking is computed from the trusted clean subset, and each pruning fraction is independently applied to the same pre-repair checkpoint before a separate 100-epoch fine-tuning branch. Thus, candidates associated with different pruning fractions belong to independent branches rather than a cumulative pruning trajectory. ANP first optimizes adversarial neuron masks and then constructs candidate models by applying different pruning thresholds to the resulting mask scores, without additional post-pruning fine-tuning. For I-BAU, candidate states correspond to models obtained after different numbers of completed outer repair rounds.

FT uses SGD with an initial learning rate of $0.01$, momentum $0.9$, weight decay $5\times10^{-4}$, and cosine learning-rate annealing. For NAD, both the teacher and student are initialized from the corresponding pre-repair model and optimized using SGD with an initial learning rate of $0.01$, momentum $0.9$, and weight decay $5\times10^{-4}$. On ResNet18-CIFAR and VGG19-BN, attention distillation is applied to the pre-classifier representation with weight $\beta_3=1000$. On PreActResNet18, attention distillation uses the last two residual stages and the global pooled representation, with $(\beta_1,\beta_2,\beta_3)=(500,1000,1000)$.

For ANP, we use $\epsilon=0.4$, one adversarial update step, and $\alpha=0.2$ during mask optimization. The neuron masks are optimized using SGD with a learning rate of 0.01 and momentum 0.9, while the adversarial noise variables are updated using SGD with a learning rate of 0.4. I-BAU uses $K=5$ fixed-point iterations per outer repair round with a fixed-point step size of 0.1. The model is updated using Adam with a learning rate of $1\times10^{-4}$ and AMSGrad, while the perturbation is optimized using SGD with a learning rate of 10. For D3, we use $\lambda=10$, $\epsilon=0.1$, and $L_p=2$, with the final classification layer selected for all evaluated architectures. D3 is optimized using SGD with a learning rate of 0.01 and momentum 0.9, together with cosine annealing over the 100-epoch optimization trajectory.

\subsection{Additional Clean-Label Validation}
\label{app:clean_label_details}

For the additional clean-label validation, we evaluate LC and SIG on CIFAR-10 using ResNet18-CIFAR, with attack target class $0$. The corresponding pre-repair models are trained for 100 epochs using SGD with an initial learning rate of $0.1$, momentum $0.9$, weight decay $5\times10^{-4}$, and a batch size of 128, with cosine annealing applied during training. In both attacks, poisoned samples are selected only from the attack-target class and retain their original labels. The poisoning rate is $5\%$ of the complete training set, corresponding to $50\%$ of the target-class training samples.

\paragraph{LC}
LC first transforms the selected target-class samples into adversarial variants using a 100-step untargeted PGD attack against a fixed robust ResNet18 surrogate. We use an $\ell_\infty$ perturbation bound of $8/255$ and a step size of $1.5/255$. A signed $3\times3$ pattern is then added at each of the four image corners using the all-corners configuration, with pattern amplitude $1.0$.

\paragraph{SIG}
SIG superimposes a horizontal sinusoidal signal $\delta\sin(2\pi jf/W)$ on the selected target-class samples, where $j$ and $W$ denote the horizontal coordinate and image width, respectively. We use $f=6$ and $\delta=40$ during both training and evaluation.

Each repair repetition uses a randomly sampled trusted clean subset containing $2\%$ of the CIFAR-10 training set. We perform three repair repetitions using seeds 1, 2, and 3, with FST and RNP using the same trusted subset within each repetition. Both repair methods use a batch size of 128. FST is optimized for 10 epochs using SGD with a learning rate of $0.01$ and $\alpha=0.2$, with candidate states collected at epochs $\{1,2,5,10\}$. RNP performs an unlearning stage for up to 20 epochs, with early stopping based on a predefined clean-accuracy threshold of $0.2$, followed by a 20-epoch recovery stage. Candidate states are constructed using recovered-mask thresholds $\{0,0.05,\ldots,0.90\}$. Representative repair states are selected using the same protocol as in Section~\ref{sec:state_selection}.

The resulting pre-repair models achieve Clean ACC and ASR of 94.72\% and 99.92\% under LC, and 94.69\% and 98.08\% under SIG, respectively.

\vspace{11pt}

\end{document}